# Efficient energy transfer in light-harvesting systems, I: optimal temperature, reorganization energy, and spatial-temporal correlations

**Jianlan Wu, Fan Liu, Young Shen, Jianshu Cao[1] and Robert J. Silbey[1]**

Department of Chemistry, Massachusetts Institute of Technology,
77 Massachusetts Ave., Cambridge, Massachusetts, 02139, USA.

E-mails: jianshu@mit.edu; silbey@mit.edu

**Abstract.** Understanding the mechanisms of efficient and robust energy transfer in light-harvesting systems provides new insights for the optimal design of artificial systems. In this paper, we use the Fenna-Matthews-Olson (FMO) protein complex and phycocyanin 645 (PC 645) to explore the general dependence on physical parameters that help maximize the efficiency and maintain its stability. With the Haken-Strobl model, the maximal energy transfer efficiency (ETE) is achieved under an intermediate optimal value of dephasing rate. To avoid the infinite temperature assumption in the Haken-Strobl model and the failure of the Redfield equation in predicting the Forster rate behavior, we use the generalized Bloch-Redfield (GBR) equation approach to correctly describe dissipative exciton dynamics and find that maximal ETE can be achieved under various physical conditions, including temperature, reorganization energy, and spatial-temporal correlations in noise. We also identify regimes of reorganization energy where the ETE changes monotonically with temperature or spatial correlation and therefore cannot be optimized with respect to these two variables.

## Contents

---

[1] To whom correspondence should be addressed.

Photosynthetic processes in plants, bacteria and marine algae provide key insights into designing artificial light harvesting systems that operate efficiently and robustly [1, 2]. The initial stages in the conversion of solar energy into chemical and other useful forms of energy for human consumption can be described by exciton dynamics with trapping and dissipation [3, 4]. Recent experimental and computational studies suggest that environmental noise can assist exciton transport and can be optimized for maximal energy transfer efficiency (ETE) [5]–[20]. Quantum entanglement in photosynthetic light-harvesting complexes has also been studied with the consideration of environmental noise [21]. In this paper, using two light-harvesting systems, FMO and PC 645, we examine the optimization of the ETE, with respect to reorganization energy, temperature, and spatial-temporal correlations.

This paper closely follows the theoretical formulation presented in a recent review [17] and further examines issues relevant for realistic light-harvesting systems. The review article addresses two questions: basic mechanisms of optimal energy transfer and systematic mapping to kinetic networks. Since environmental noise helps maximize energy transfer efficiency, it stands to reason that light-harvesting systems can be optimized to achieve best performance under a given environment, which leads to the proposal of optimal design. Previous work on simple models [10, 17] and FMO [13]-[16] use the Haken-Strobl model, an infinite temperature model, and therefore report optimization a function of single parameter, the pure dephasing rate. It remains an open question if the ETE can be optimized as a function of temperature, reorganization energy, bath correlation time, and spatial correlation. In this paper, we will further examine the idea of noise-enhanced optimal energy transfer with explicit considerations of different descriptions of dissipation, general parameter variations, and spatial-temporal correlations. We will show that ETE can be optimized with general parameter-dependence and that spatial correlation and thermal fluctuations can either enhance or suppress the ETE. We have also formulated a conceptual framework to understand noise-assisted exciton trapping and will discuss the theory elsewhere [18].

This paper is the first of a two-part series, with paper I on optimization of ETE and paper II on mechanisms and network kinetics. Though intrinsically quantum mechanical, exciton dynamics are often described by site-to-site hopping as in random walk [22], and quantum coherence is often related to delocalized tunneling [23]-[25]. The kinetic expansion unifies coherent tunneling and incoherent hopping by systematically reducing exciton dynamics to network kinetics [17]. In paper II, this approach will be applied to analyze network structures of light-harvesting systems and to predict the contribution of quantum coherence [26].

The paper is organized as follows: In section 1, we introduce the exciton dynamics model and define the ETE for light-harvesting energy transfer. In section 2, we apply the Haken-Strobl model to investigate the optimal ETE in FMO as a function of the dephasing rate. In contrast to earlier studies, our study emphasizes the initial condition dependence, the approximation of the efficiency with the average trapping time, and the secular approximation. In section 3, we apply the generalized Bloch-Redfield (GBR) equation to explore the detailed energy transfer optimization conditions in FMO with reorganization energy, temperature, and spatial-temporal correlations. In section 4, we calculate the optimal ETE in another light-harvesting system, PC 645 with the GBR approach. The general optimization feature of the energy transfer process is illustrated by our studies. In section 5, we discuss our conclusions.

1. **Exciton dynamics in light-harvesting systems and energy transfer efficiency**

When sunlight shines on light-harvesting pigments, an absorbed photon can excite the photosynthetic system. The excitation energy is then transported through an energy transfer network to the reaction center for subsequent charge separation, resulting in energy trapping. During transport, the energy can decay when excitation energy is lost as heat via irreversible electron-hole recombination and can be re-distributed through interaction with protein environments. Hence, the exciton dynamics for the light-harvesting system follows the Liouville equation [27]-[29], [17],

$$\partial_t \rho = -L\rho = -[L_{sys} + L_{trap} + L_{decay} + L_{dissp}]\rho \qquad (1)$$

where $\rho$ is the reduced density matrix of the exciton system, and each term of the Liouville superoperator $L$ describes a distinct dynamic process.

The quantum coherent evolution, $L_{sys}\rho = i[H,\rho]/\hbar$, is controlled by the system Hamiltonian, which is given by $H_{nm} = \delta_{n,m}\varepsilon_n + (1-\delta_{n,m})J_{nm}$ from the tight-binding model in the local site basis set representation [28]. The diagonal elements, $\varepsilon_n$, define the site energies, whereas the off-diagonal elements, $J_{nm} = J^*_{nm}$, define the dipole-dipole interaction coupling strength between two distinct sites.

The two irreversible energy decay channels are exciton decay and trapping. The electron-hole recombination process is described by the decay term $[L_{decay}]_{nm} = (k_{d;m} + k_{d;n})/2$, where $k_{d;m}$ is the decay rate at site $m$, and $L_{nm}$ represents the diagonal element of the Liouville operator, $L_{nm} = L_{nm,nm}$. Similarly, localization at the charge separation state is described by the exciton trapping term, $[L_{trap}]_{nm} = (k_{t;m} + k_{t;n})/2$, where $k_{t;m}$ is the trapping rate at site $m$.

The system-bath interaction, $H_{SB} = \sum_m Q_m B_m$, is applied to describe exciton dissipative dynamics, where $Q_m = |m><m|$ and $B_m$ are system and bath quantum operators, respectively [27]-[29]. The bath influence is characterized by the time-correlation function, $C_{mn}(t) = <B_m(t)B_n>$, which is related to the spectral density $J_{mn}(\omega)$ by $C_{mn}(t) = \int_0^\infty [\coth(\hbar\beta\omega/2)\cos(\omega t) - i\sin(\omega t)]J_{mn}(\omega)d\omega$. Alternatively, we can introduce a time-dependent site energy fluctuation $\delta\varepsilon_m(t)$. In the Haken-Strobl model, a spatially-uncorrelated classical white noise follows $<\delta\varepsilon_m(t)> = 0$ and $<\delta\varepsilon_m(t)\delta\varepsilon_n> = \Gamma^*\delta(t)\delta_{m,n}$, which is valid in the infinite temperature limit [30]. In the local site basis set representation, dissipation becomes simply decoherence,

$$[L_{dissp}]_{nm} = (1-\delta_{n,m})\Gamma^*, \qquad (2)$$

and exciton dynamics can be solved exactly by the second order expansion form [23, 30].

At finite temperatures, a theoretical framework of quantum dissipative dynamics differing from the Haken-Strobl model is required to include detailed balance and the memory effect in slow bath relaxation. In the weak dissipation regime, the Redfield equation (with or without the secular approximation) [31] has been widely used in modeling exciton dynamics, e.g., several previous studies of exploring energy transfer efficiency (ETE) [32, 33]. However, this approach can lead to unphysical predictions in the strong dissipation regime. With the introduction of auxiliary fields, exciton dyanamics can be studied in the generalized Bloch-Redfield (GBR) equation approach, which is more reliable over a broad regime of dissipation and predicts the correct strong dissipation limit [34].

In this paper, we first use the Haken-Strobl model to study the optimization of the exciton energy transfer (EET), obtaining a simple physical picture. We will then use the GBR approach with the Debye spectral density to explore optimal EET conditions with various control parameters, including temperature, reorganization energy, and spatial-temporal correlations of bath.

The energy trapped at the reaction center, i.e., the rate process described by $L_{trap}$, represents effective energy transfer, whereas the rate process described by $L_{decay}$, represents ineffective energy loss. The energy transfer efficiency is defined by the branching ratio of the energy trapping process, [4, 13, 32, 35], i.e.,

$$q = \text{Tr} \int_0^\infty L_{trap} \rho(t) dt = \frac{\sum_n k_{l,n} \tau_n}{\sum_n k_{l,n} \tau_n + \sum_n k_{d,n} \tau_n} , \qquad (3)$$

where the trace denotes the sum of diagonal population elements of the matrix. This definition can be simplified to $q = \sum_n k_{t,n} \tau_n$ since the total depletion probability in the denominator is normalized to one. The mean residence time at each site of the exciton system is $\tau_n = \int_0^\infty dt \rho_n(t)$, where the population $\rho_n$ is represented as the diagonal elements on the density matrix $\rho_n = \rho_{n,n}$. To simplify, we assume that $k_d$ is identical at each site together with a necessary condition $k_t \gg k_d$ for the high-efficient excitation energy transfer (EET) network. Then, the $k_d$-dependence of the residence time becomes negligible, giving

$$q \approx \frac{1}{1 + k_d <t>} \qquad (4)$$

Here, the residence time is approximately $\tau(k_d) \approx \tau(0)$ with the normalization condition $\sum_n k_{t,n} \tau_n (k_d = 0) = 1$, and $\langle t \rangle = \sum_n \tau_n (k_d = 0)$ is the mean first passage time to the trap state in the absence of decay (i.e., the average trapping time). As shown later in this paper, the optimal ETE is determined reliably by the minimal trapping time. The above definitions follow closely those introduced in reference [17]. Using the stationary solution to equation (1), $L \int_0^\infty \rho(t) dt = L\tau = \rho(0)$, we obtain the average trapping time as

$$<t> = \text{Tr}[(L_{sys} + L_{trap} + L_{dissip})^{-1} \rho(0)] \qquad (5)$$

where $\rho(0)$ is the initial condition.

**2. Optimization in the Haken-Strobl model of FMO**

As the first light-harvesting system in this paper, we consider the photosynthetic Fenna-Matthews-Olson (FMO) protein complex in green sulfur bacteria with seven bacteriochlorophyll (BChl) sites [5]-[7], [36]-[41]. The experimental Hamiltonian has the following matrix elements (in cm$^{-1}$) [38, 39]:

$$H_{BChl_m,BChl_n} = \begin{bmatrix} 280 & -106 & 8 & -5 & 6 & -8 & -4 \\ -106 & 420 & 28 & 6 & 2 & 13 & 1 \\ 8 & 28 & 0 & -62 & -1 & -9 & 17 \\ -5 & 6 & -62 & 175 & -70 & -19 & -57 \\ 6 & 2 & -1 & -70 & 320 & 40 & -2 \\ -8 & 13 & -9 & -19 & 40 & 360 & 32 \\ -4 & 1 & 17 & -57 & -2 & 32 & 260 \end{bmatrix}. \quad (6)$$

The site closest to the reaction center, BChl 3, is designated as the only site with a nonzero trapping rate $k_t = 1$ ps$^{-1}$ [13-16] To compare with real experiments, the initial population is considered to be at either BChl 1 or BChl 6, i.e., $\rho_{m=1}(0)=1$ or $\rho_{m=6}(0)=1$.

For optimal energy transfer efficiency, the exciton dynamics of the FMO complex uses environmental noise to increase the transfer efficiency [13]-[16], previously studied using the Haken-Strobl model. In this section, we will revisit this simple model but emphasize three new aspects of the EET process in FMO: (i) the effect of different initial conditions, (ii) verification of approximating the energy transfer efficiency with the average trapping time, and (iii) limitation of the secular approximation in the Redfield equation.

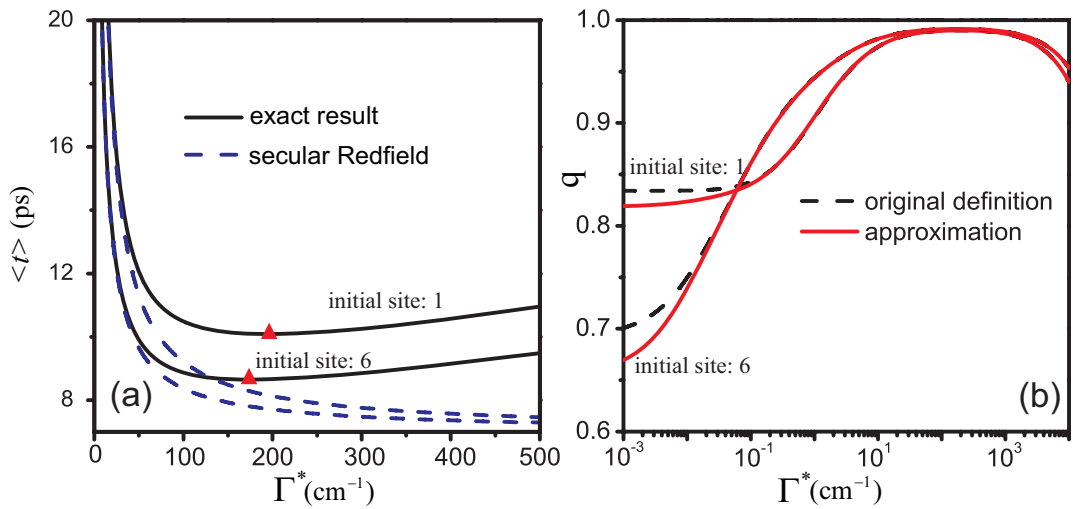

**Figure 1.** The calculation of the Haken-Strobl model in FMO under two initial conditions ($\rho_{n=1}(0)=1$ and $\rho_{n=6}(0)=1$, as indicated by labels in the figures). (a) The dependence of the average trapping time $<t>$ on the pure dephasing rate $\Gamma^*$: The solid lines are the exact results, and the dashed lines are the results calculated from the secular Redfield equation in the exciton basis set representation. The triangles denote optimal conditions where $<t>$ is minimal. (b) The comparison of two methods for calculating the ETE $q$: the dashed lines are from the original

definition of $q$ in equation (3) and the solid lines are the exact results from the approximated form of $q$ in equation (4). Here the trapping rate is $k_t = 1$ ps$^{-1}$ and the decay rate is $k_d = 1$ ns$^{-1}$.

We calculate the average transfer time $<t>$ as a function of the pure dephasing rate $\Gamma^*$ with the two different initial conditions. As shown in figure 1(a), a characteristic optimization phenomenon of $<t>$ is observed: At dephasing rates less than 50 cm$^{-1}$, the average trapping time has a steep increase as $\Gamma^*$ decreases; at dephasing rates more than 200 cm$^{-1}$, the average time marginally increases with $\Gamma^*$. The interplay between coherent population oscillation and dissipative population re-distribution leads to an optimal transfer efficiency at an intermediate dephasing rate. The explicit optimal values are $<t>_{min} = 10.1$ ps with $\Gamma^*_{opt} = 195$ cm$^{-1}$ and the initial condition defined at BChl 1, and $<t>_{min} = 8.65$ ps with $\Gamma^*_{opt} = 175$ cm$^{-1}$ and the initial condition defined at BChl 6. Different initial conditions result in similar but distinguishable results, which can be interpreted by a simple physical argument. In FMO, there are two dominating EET pathways: $1 \to 2 \to 3$ and $6 \to (5,7) \to 4 \to 3$, where each number represents a specific BChl site [40]. With fewer sites involved, the first pathway exhibits stronger quantum coherence than the second pathway. To compensate the inefficient energy transfer in the fully coherent limit, a larger optimal dephasing rate is needed for the first pathway than the second pathway, i.e., $\Gamma^*_{opt}(\text{Bchl}\,1) > \Gamma^*_{opt}(\text{Bchl}\,6)$.

The physical quantity of interest is quantum efficiency, whereas the theoretical quantity computed in this and our preceding papers is the average trapping time. The approximate expression in (4) establishes the connection between these two quantities. We examine the validity of this approximation by evaluating the ETE in FMO with the exciton recombination rate $k_d = 1$ ns$^{-1}$ [13-16]. As shown in figure 1(b), the average trapping time provides a reliable measurement for the ETE as $\Gamma^*$ changes by five orders of magnitude ($0.1$ cm$^{-1} < \Gamma^* < 10^4$ cm$^{-1}$), except for a slight underestimation in the weak dephasing limit ($\Gamma^* < 0.1$ cm$^{-1}$). Thus, we will focus on the calculation of $<t>$ in the rest of the paper.

We now examine the failure of the secular approximation in the intermediate to strong damping regime. As demonstrated in our forthcoming paper, a physically reasonable description of dissipation in energy transfer networks will always lead to optimal noise for the maximal efficiency [18]. Here the optimal noise refers a finite value of noise where the ETE reaches the maximal. We present here a counter-example, where the secular approximations adopted in the Redfield equation leads to an unphysical prediction in the average trapping time. The Haken-Strobl model introduces classical Gaussian fluctuations in the site energy and therefore is rigorously described by the dephasing operator in (2) in the local basis set [23, 30]. A unitary basis set transformation of the Liouville equation in (1) does not change the exciton dynamics and will give exactly the same prediction, which is confirmed by our calculation using the full Redfied equation in the eigen-exciton basis set [31]. In contrast, as shown in figure 1 (a), an additional secular approximation in the Redfield equation leads to a plateau of the average trapping time as the dephasing rate increases. In the secular Redfield equation, the relaxation rate is proportional to $\Gamma^*$, suggesting instantaneous population transfer to the trap state (e.g., BChl 3 in FMO) for large $\Gamma^*$. Therefore, the plateau in figure 1 is given by the trapping rate, i.e., $\langle t \rangle \propto 1/\Gamma^* + 1/k_t$. However, strong dephasing should always lead to classical behavior, i.e., hopping kinetics, which results in a slow-down in the diffusion to the trap state. As in the diffusion-controlled reaction, diffusion becomes the rate-limiting step in the intermediate to strong dissipative regime, giving $\langle t \rangle \propto \Gamma^*$. The failure of the secular approximation has been discussed in several contexts before [42, 43], and is now used to explain the observed discrepancy between the Haken-Strobl model calculation and secular Redfield calculation (i.e. the quantum jump method). Although our discussion here is limited to the Haken-Strobl model defined for infinite temperature, we show in the next section that the conclusion about the secular approximation applies remains valid for exciton dynamics at finite temperatures.

Noise-assisted energy transfer using the Haken-Strobl model has been reported recently by several groups [10], [13]-[16], but a basic question still remain to be answered: Is the optimal dephasing rate observed in FMO a general rule for realistic light-harvesting systems or a special case obtained from model exciton systems? Without including temperature and memory effects, predictions based on the Haken-Strobl model may not be relevant under realistic conditions. Although previous studies have attempted to include temperature dependence and realistic spectral densities, the theoretical methods involved can lead to qualitatively unphysical predictions in the intermediate and strong dissipation regimes [32, 33]. In the next section, our study at finite temperatures will demonstrate the general optimal conditions in FMO, including temperature, reorganization energy, and spatial-temporal correlation.

## 3. Optimization in the generalized Bloch-Redfield equation description of FMO

### 3.1. Generalized Bloch-Redfield equation

In the previous section, we used the Haken-Strobl model, defined in the infinite temperature limit, to illustrate the possibility of optimization in the EET process. At finite temperatures, forward and backward energy transfer rates need to satisfy the detailed balance condition. In the weak dissipation regime, the Redfield approach is able to reliably describe exciton dissipative dynamics. With the increase of system-bath coupling and the slow-down of bath relaxation, more expensive methods such as the hierarchic expansion for the Gaussian bath are required to describe exciton dynamics accurately [44]-[47]. Although the hierarchic approach has been applied to FMO, a rigorous investigation of quantum dissipative dynamics in large-scale exciton systems can be numerically difficult. To capture the relevant optimization feature of EET dynamics in a qualitatively reliable way, we will apply a generalized Bloch-Redfield (GBR) equation approach, derived from the second-order cumulant expansion. For many bath spectral densities such as Debye or Ohmic, the corresponding time correlation functions can be (numerically) expanded using exponential functions [46],

$$C_{mn}(t) = c_{mn} \sum_{i=0}^{\infty} (f_i^r + i f_i^i) e^{-\nu_i t}, \qquad (7)$$

where $c_{mn}$ is the spatial correlation coefficient between sites $m$ and $n$, $\nu_i$ is the relaxation rate of the $i$-th bath mode, and $f_i^r$ ($f_i^i$) is the real (imaginary) part of the expansion coefficient. In the frequency domain, the expansion is applied to both spectral density $J(\omega)$ and temperature dependence $\coth(\hbar\omega/2k_B T)$. With the facilitation of auxiliary fields $g_{m;i}(t)$ at site $m$, the GBR equation for exciton dynamics is written as

$$\dot{\rho}(t) = -(L_{sys} + L_{trap})\rho(t) - i\sum_{i=0}^{\infty}\sum_{m}[Q_m, g_{m;i}(t)]$$

$$\dot{g}_{m;i}(t) = -(L_{sys} + L_{trap} + \nu_i)g_{m.i}(t) - if_i^r[Q_m, \rho(t)] + f_i^i[Q_m, \rho(t)]_+ \quad (8)$$

where $Q_m = |m\rangle\langle m|$ is the system operator together with $Q_m = \sum_n c_{mn}Q_n$, and $[A,B]_+ = AB + BA$ is the anti-commutator. The initial value of $g_{m;i}(t)$ is zero. Equation (8) is a generalization of its original form [34]. Here the memory effect in the dissipative dyanamics due to the interaction with bath is represented by the auxiliary field, which can be considered as elements of the projection operator $Q$ ($I$-$P$) in the Nakajima-Zwanzig projection operator technique.

In this paper, we will use the spatially-correlated Debye spectral density, $J_{mn}(\omega) = c_{mn}J(\omega)$ and $J(\omega) = (2\hbar/\pi)\lambda\omega D/(\omega^2 + D^2)$, where the reorganization energy $\lambda$ represents the system-bath coupling strength and the Debye frequency $D$ is the bath relaxation rate (the inversion of bath temporal correlation). The choice of spectral density will not affect our general conclusion. Following the Matsubara expansion [27], the bath time correlation function can be written explicitly as the exponential expansion in (7), where $\nu_i = 2\pi i k_B T/\hbar$ is the $i$-th ($i > 0$) Matsbura frequency and $\nu_0 = D$. Dissipation is thus characterized by four parameters: reorganization energy $\lambda$, temperature $T$, bath relaxation rate $D$, and bath spatial correlation $c_{mn}$.

Due to the equivalence of the generalized Bloch-Redfield equation and the second-order time-nonlocal expansion form, the trapping time can be alternatively calculated by the Laplace transform in the eigen-exciton basis set using spectral density. The results from two approaches are exactly the same, which demonstrates the basis set invariance of the GBR approach for dissipative dynamics in EET systems.

*3.2. Optimization with reorganization energy*

We begin with studying the influence of reorganization energy on the EET process. As an example, we assume zero spatial correlation $c_{mn} = \delta_{m,n}$, an experimentally reasonable temporal correlation $D^{-1} = 50$ fs, and room temperature $T = 300$ K. The trapping time calculated from the GBR approach is plotted in figure 2(a). As $\lambda$ increases, the trapping time quickly decreases from the fully coherent value by two orders of magnitude to the minimal value, corresponding to the optimal ETE, and then slowly increases again. The explicit optimal values are $<t>_{min} = 3.39$ ps with $\lambda_{opt} = 92$ cm$^{-1}$ for the initial population at BChl 1 and $<t>_{min} = 3.27$ ps with $\lambda_{opt} = 73$ cm$^{-1}$ for the initial population at BChl 6, respectively. These results including the initial condition dependence are similar to those derived in the Haken-Strobl model, except that the reorganization energy $\lambda$ is used instead of the dephasing rate $\Gamma^*$ (see figure 1(a)). At high temperatures, the dephasing rate can be effectively written as

$$\Gamma^* \approx 4k_B T\lambda/D, \quad (9)$$

which is the zero-frequency limit of $J(\omega)\coth(\hbar\omega/2k_BT)$. The linear relation between $\Gamma^*$ and $\lambda$ indicates that these two parameters behave qualitatively in the same way for the EET process. As we will find later, equation (9) can also be used to interpret the optimization behavior with other control parameters.

In figure 2(b), we compare the predictions of $<t>$ from several Master equation approaches. Both the full Redfield equation and its secular form lead to an unphysical plateau of $<t>$ for large $\lambda$. This plateau was observed in recent calculations of FMO [32, 33] and, as discussed in the previous section, is due to the time-local approximation of the Redfield equation. In the third approach, we only keep the real part of Redfield tensors, which is equivalent to the Haken-Strobl model but with temporally-correlated noise. This extended Haken-Strobl model recovers the optimization behavior but overestimates noise enhancement: the predicted $\lambda_{opt}$ is 50% smaller whereas $<t>_{min}$ is 50% larger than those from the GBR approach. These observations are consistent with an early study by Ishizaki and Fleming in the two-site model system [45, 46]. On the other hand, with a time-nonlocal form for dissipative dynamics, our GBR approach is qualitatively accurate in calculating the variation of $<t>$ with $\lambda$. A comparison with the sophisticated hierarchic approach will be necessary to test the quantitative accuracy of results in this paper. However, from the practical point of view, the GBR serves as a simple but unifying approach to investigate exciton dynamics in a broad range of parameter space with low computational cost.

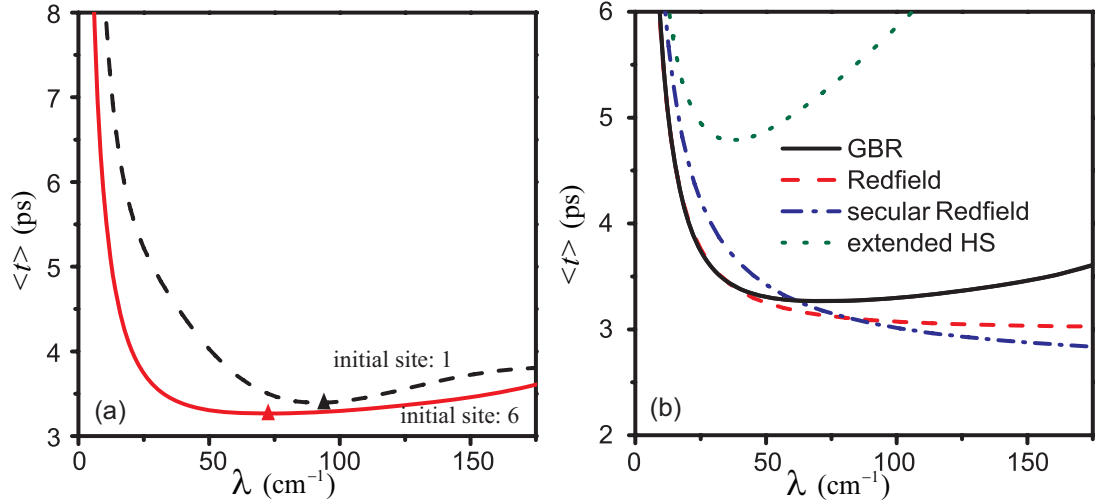

**Figure 2.** (a) The calculation of trapping time vs. the reorganization energy in FMO using the generalized Bloch-Redfield (GBR) equation with the Debye spectral density under the two initial conditions labeled in the figure. The parameters used are shown in text. The triangles denote optimal conditions. (b) The comparison of the GBR approach and three approaches based on the Redfield equation. Only the initial site condition defined at BChl 6 is used for the calculation. The solid line is from the GBR equation (the same as the solid line in the (a)); the dashed line is from the full Redfield equation; the dotted-dashed line is from the secular Redfield equation; and the dotted line from the extended Haken-Strobl model (see text).

*3.3. Optimization with temperature*

Next we study the dependence of the trapping time on temperature. With the same spatial-temporal correlation ($c_{mn} = \delta_{m,n}$ and $D^{-1} = 50$ fs) used in the previous subsection, we fix the reorganization energy at $\lambda = 35$ cm$^{-1}$, which is a physically reasonable estimation from experiments [47]. The results under the two initial conditions are reported in figure **3**. With the initial condition at BChl 1, the minimal trapping time is $<t>_{min} = 4.2$ ps at optimal temperature $T_{opt} = 162$ K, but the increase of $<t>$ with decreasing $T$ at low temperatures is weak. With the initial condition at Bchl 6, the trapping time monotonically increases with temperature, and the maximum ETE is at zero temperature. Since the trapping time is kept in the order of ps, FMO shows a strong robustness against the change of temperature.

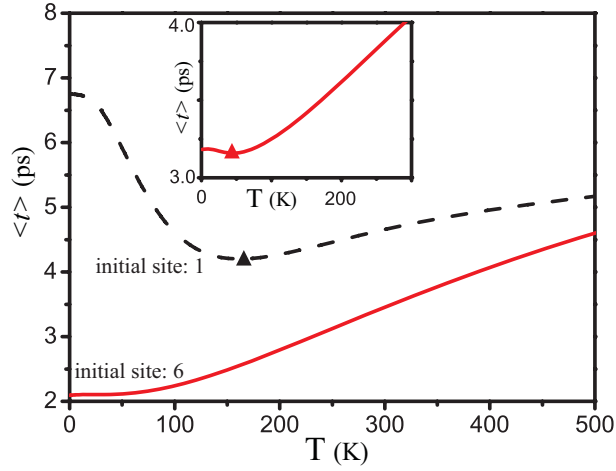

**Figure 3.** The calculation of trapping time vs. temperature in FMO using the GBR approach with the Debye spectral density under the two initial conditions labeled in the figure. The reorganization energy is $\lambda = 35$ cm$^{-1}$ and the other parameters used are shown in text. The inset shows the result with another reorganization energy $\lambda = 20$ cm$^{-1}$ under the initial condition at BChl 6. The triangles denote optimal conditions for the ETE is maximized at finite temperatures.

In dissipative dynamics, the exact temperature dependence is determined by $\coth(\hbar\omega/2k_BT)$, which changes from a linear function, $2k_BT/\hbar\omega$, at high temperatures to a constant at low temperatures. Non-vanishing dissipation is needed to maintain the equilibrium energy distribution even at $T \to 0$. Because dissipation always increases with temperature for a given spectral density, the zero-temperature limit of dissipation determines the general behavior of $T$-dependence of the trapping time and the ETE. When the zero-temperature limit of dissipation is weaker than the optimal value, increasing temperature can increase the dissipation strength and therefore help the system to achieve the maximum ETE at a finite temperature. When the zero-temperature limit of dissipation is stronger than the optimal value, the trapping time increases monotonically with temperature and the system will be further away from the optimal condition. Hence, depending on the reorganization energy, temperature can behave in two different ways in the EET process. To verify this prediction, we use a small reorganization energy of $\lambda = 20$ cm$^{-1}$, and start the initial population at BChl6. The inset of figure 3 represents an optimization curve where $T_{opt} = 46$ K is observed.

*3.4. Optimization with temporal correlation*

In the Markovian approximation for dissipative dynamics, bath relaxation is assumed to be much faster than system dynamics. This assumption may not be reliable in light-harvesting systems due to the complexity of the protein scaffold. The time scale of bath relaxation ($D^{-1}$) in FMO is usually in the orders of 10-100 fs [47], smaller but comparable to that of exciton dynamics. The non-Markovian memory effect can enhance quantum coherence up to 500 fs as observed experimentally [5, 7]. As a time-integrated effect of exciton dynamics, energy transfer efficiency can depend strongly on the temporal correlation of bath, which is shown by the $D$ dependence of the effective dephasing rate at high temperatures in equation (9).

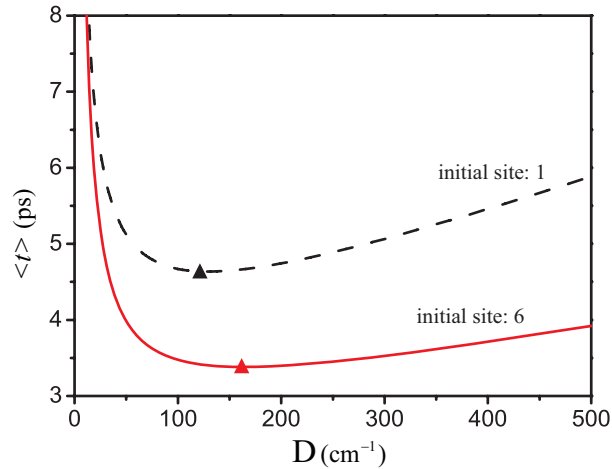

**Figure 4.** The calculation of trapping time vs. the bath relaxation rate in FMO using the GBR approach with the Debye spectral density under the two initial conditions labeled in the figure. The parameters used are shown in text. The triangles denote optimal conditions.

Here we apply zero spatial correlation $c_{mn} = \delta_{m,n}$, reorganization energy $\lambda = 35$ cm$^{-1}$, and room temperature $T = 300$ K. Under the two initial conditions, the trapping time $<t>$ is calculated as a function of the Debye frequency $D$ in the GBR approach. As shown in figure 4, the temporal correlation of bath leads to the optimal efficiency at an intermediate level of $D$. In our calculation, optimal conditions are $<t>_{min} = 4.48$ ps with $D^{-1}_{opt} = 32$ fs and the initial condition at BChl 1 and $<t>_{min} = 3.41$ ps with $D^{-1}_{opt} = 41$ fs and the initial condition at BChl 6. Qualitatively, the influence of the Debye frequency on the EET process can be interpreted by the inverse relation between the effective dephasing rate $\Gamma^*$ and $D$ in equation (9). Our results further suggest that a more quantitative characterization of bath temporal correlation, rather than being labeled as Markovian or non-Markovian, is necessary in studying the EET process.

*3.5. Optimization with spatial correlation*

Recent experimental and theoretical studies have suggested strong spatial correlation can be relevant in exciton dynamics in light –harvesting systems[6]. Spatial correlation can be quantified through different approaches. Here, we use an exponentially decaying function [33, 41, 48],

$$c_{mn} = \exp(-R_{mn}/R_0), \qquad (10)$$

to define the spatial correlation with the site-site distance $R_{mn}$ and the correlation length $R_0$ [6]. With the assumption that $R_0$ in FMO is the same as that in the reaction center, we rewrite equation (10) as

$$c_{mn} = c^{R_{mn}/R_{RC}} \qquad (11)$$

where $c = 0.9$ and $R_{RC} = 11.93$ Å are the coefficient of the spatial correlation and the site-site distance in the reaction center, respectively. Instead of the reported value for $c$, we use $c$ as the free parameter, ranging from 0 to 1, to characterize spatial correlation

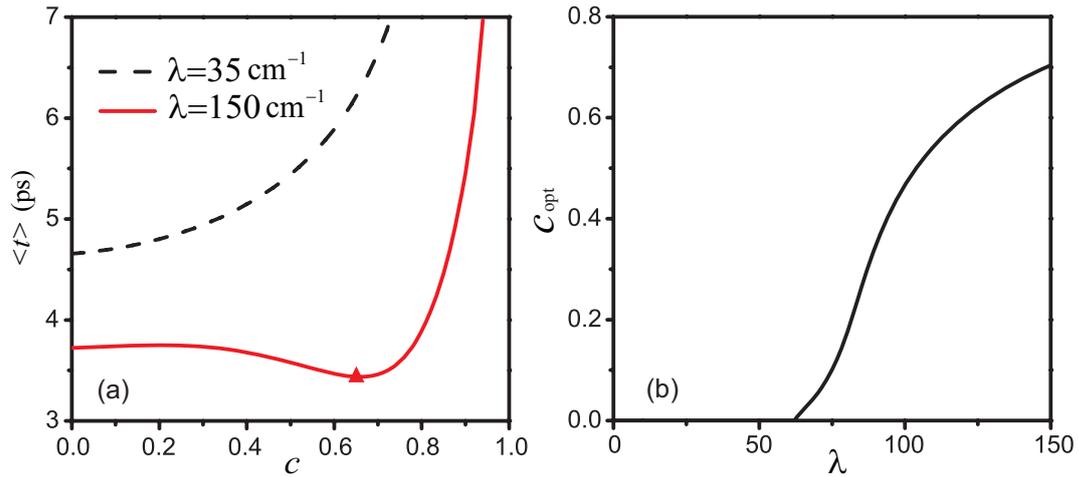

**Figure 5.** (a) The calculation of trapping time vs. the spatial correlation strength in FMO using the GBR approach with the Debye spectral density under the initial condition defined at BChl 1. The reorganization energies are $\lambda = 35$ cm$^{-1}$ for the dashed line, and $\lambda = 150$ cm$^{-1}$ for the solid line. The other parameters used are shown in text. The triangle denotes the optimal condition where the ETE is maximized at a nonzero value of $c$. (b) The optimal spatial correlation for the minimal trapping time vs. the reorganization energy under the initial condition at BChl 1.

To compare with experiments, we use realistic parameters: reorganization energy $\lambda = 35$ cm$^{-1}$, temporal correlation $D^{-1} = 50$ fs, and room temperature $T = 300$ K, to calculate the dependence of the trapping time $<t>$ on the spatial correlation coefficient $c$. For simplicity, we present the result with the initial population localized at BChl 1 (the result with BChl 6 as the initial condition is similar). As shown in figure 5(a), the average trapping time monotonically increases with spatial correlation. The suppression of transfer efficiency due to spatial correlation has been shown in an early numerical simulation [48]. However, when we use a much larger reorganization energy, $\lambda = 150$ cm$^{-1}$, the EET process reaches the maximal efficiency with an intermediate level of spatial correlation, $c_{opt} = 0.66$. To understand these two different behaviors, we notice that a positive spatial correlation always reduces the dissipation strength. For example, for a quantum two-level system, the energy fluctuations between the two levels can be characterized by an effective reorganization energy defined as $\lambda_{eff} = \lambda(1-c)$. In the completely correlated case ($c = 1$), quantum coherence between these two levels can persist under an arbitrarily strong energy fluctuation ($\lambda_{eff} = 0$). Therefore, when the reorganization energy is smaller than the optimal value, spatial correlation further reduces the effective reorganization energy and there is no optimal ETE as a function of $c$, i.e., the ETE is maximized at $c = 0$. In contrast, when the reorganization energy is larger than the optimal value, the ETE can be maximized at a nonzero $c$ by the reduced dissipation at an optimal value of spatial correlation. Indeed, as shown in figure 5(b), the nonzero optimal spatial correlation starts to appear for $\lambda > 60$ cm$^{-1}$ with the initial condition at BChl 1.

## 4. Optimization in PC645

In this section, we consider another light-harvesting system, phycocyanin 645 (PC645), in cryptophyte algae. Exciton dynamics of PC645 can be studied using an effective eight-site Hamiltonian model, where the excitation energy is transferred from a center dimer dihyrobiliverdin (DBV) β50/β61 with the highest energy to two phycocyanins (PCB) β82 with the lowest energy during the initial step [8, 49, 50]. To compare with experiments where the initial state deviates from an eigenstate [8], we apply an incoherent initial condition at two DBV bilins with equal population, and the trapping rates $k_t = 1$ ps$^{-1}$ at two PCB 82 bilins. Hence, we limit our investigation to the initial step from DBV bilins to PCB bilins within the timescale of ps.

As in the study in the previous section, the GBR approach is applied to explore the optimal ETE and its dependence on various control parameters. Experimentally, the bath spectral density is approximated by a two-frequency Debye model with $D_1^{-1} = 50$ fs and $D_2^{-1} = 1.5$ ps [50]. The reorganization energies for these two frequencies are chosen to be the same, giving $J_{mn}(\omega) = c_{mn}[J_1(\omega) + J_2(\omega)]$ with $J_{i=1,2}(\omega) = (2\hbar/\pi)\lambda\omega D_i/(\omega^2 + D_i^2)$. Here we adopt the Gaussian bath model with these two bath time scales and investigate the influence of varying reorganization energy, temperature, and spatial correlation on the EET process.

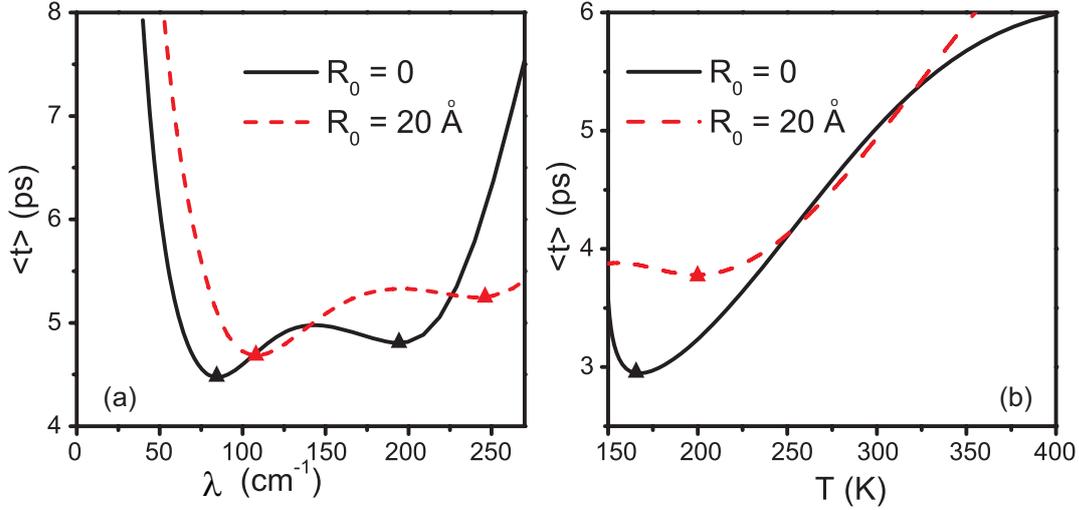

**Figure 6.** The calculation of trapping time vs. (a) reorganization energy and (b) temperature in PC 645 using the GBR approach with the two-frequency Debye spectral density. The parameters used are shown in text. The solid line corresponds to the zero spatial correlation whereas the dashed line corresponds to the correlation length of $R_0 = 20$ Å. The triangles denote optimal conditions.

Without spatial correlation, we calculate the trapping time $<t>$ as a function of reorganization energy $\lambda$ at room temperature $T = 300$ K; the result is plotted as a solid line in figure 6(a). Similar to the case in FMO, the trapping time is large in the regimes of small and large reorganization energies, but the reorganization energy-dependence of ETE at intermediate $\lambda$ becomes complicated. Two local minima appear in the trapping time due to the fact that the two Debye frequencies are highly separated in time scales. The explicit values of these two local minima are $<t>_{min} = 4.53$ ps with $\lambda_{opt} = 83$ cm$^{-1}$ and $<t>_{min} = 4.84$ ps with $\lambda_{opt} = 193$ cm$^{-1}$. Thus, the optimization behavior is also dependent on the detailed structure of bath. In comparison, we also present the result with consideration of spatial correlation. Following the original exponentially decaying function in equation (10), the spatial correlation is characterized by the correlation length $R_0$. The result of the trapping time at $R_0 = 20$ Å is plotted as a dashed line in figure 6(a), which also exhibits a range of intermediate $\lambda$ for minimized trapping time and optimized ETE, but the values of $\lambda_{opt}$ become larger than those without spatial correlation. The increase of $\lambda_{opt}$ arises from the reduction of dissipation strength due to spatial correlation.

Using the reorganization energy, $\lambda = 135$ cm$^{-1}$ [50], we next calculate the trapping time as a function of temperature, with and without spatial correlation. As shown in figure 6(b), $<t>$ can be minimized at an intermediate temperature: $<t>_{min} = 2.95$ ps with $T = 167$ K and zero spatial correlation, and $<t>_{min} = 3.78$ ps with $T = 199$ K and $R_0 = 20$ Å. The GBR approach in PC 645 becomes less reliable at low temperatures due to much larger reorganization energy than that in FMO. Our study is thus limited to a physically acceptable regime of the GBR approach with $T > 150$ K, below which the inversion of the Liouville superoperator may become unreliable. The value of the cutoff temperature, 150 K, is chosen for numerical convenience. .

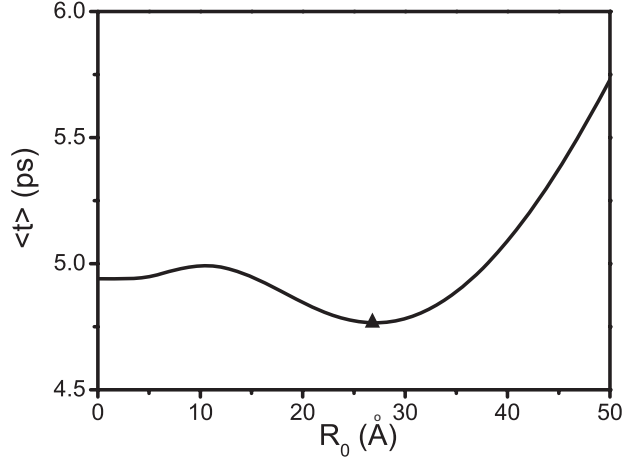

**Figure 7.** The calculation of trapping time vs. correlation length in PC 645 using the GBR approach with the two-frequency Debye spectral density. The parameters used are shown in text. The triangle denotes the optimal condition.

In our final calculation, we study the dependence of the trapping time on spatial correlation (using the correlation length). With temperature $T = 300$ K and reorganization energy $\lambda = 135$ cm$^{-1}$, the result is plotted in figure 7. The maximum ETE is observed with the minimum trapping time at $R_{0;opt} = 27$ Å. Since the interaction between chromophores of PC 645 and protein backbone is covalent, the reorganization energy is much larger than that of FMO and the optimization on spatial correlation becomes more pronounced, as discussed in the previous section. In addition, the dependence of the trapping time on spatial correlation becomes complicated with a local maximum trapping time appearing around $R_0 = 10.4$ Å.

## 6. Conclusions and discussions

This paper is built on recent efforts in exploring quantum effects in light-harvesting systems for the purpose of optimizing energy transfer efficiency (ETE). Employing the analytical methods and concepts introduced in a recent review article [17], our comprehensive study in FMO and PC 645 with both the Haken-Strobl model and the generalized Bloch-Redfield (GBR) equation reveals a general feature of the EET process: *the interplay of coherent dynamics and environmental noise leads to the optimal ETE at an intermediate level for various variables.* Explicitly, the reorganization energy and the bath relaxation rate, i.e., Debye frequency, yield non-monotonic-dependence and thus lead to the optimal ETE. On the other hand, a nonzero value of temperature can optimize the ETE only when the dissipation is weak (less than its optimal value); a nonzero value of spatial correlation can optimize the ETE only when the dissipation is strong (large reorganization energy). If the dissipation strength does not fall in the chosen regime, temperature and spatial correlation can lead to monotonic changes in ETE and we will not observe the optimal ETE. This observation clearly demonstrates a difference between spatial correlation and temporal correlation, though both correlations can enhance quantum coherence. Although the findings in this paper are reported for FMO and PC 645, the underlying principles should be general. In forthcoming paper, we will identify the exciton states orthogonal to the trap state as the sufficient condition for non-zero optimal noise [18], which extends the invariant subspace proposed by Plenio *et. al.* extracted from a specific fully-connected network model [15]. The orthogonal exciton subspace provides a unifying framework to explain optimal noise, initial preparation, coherent phase modulation, and spatial-temporal correlations.

For light-harvesting systems, parameters under realistic situations are often not the same as their optimal values shown in our calculation. However, light-harvesting systems exhibit robustness against the change of variables since the resulting trapping time varies slowly and the time scales of energy trapping and decay are highly separated. For example, the variation of the trapping time at room temperature in FMO is within one piecosecond as the reorganization energy changes from 50 cm$^{-1}$ to 150 cm$^{-1}$. A physical interpretation will be found based on kinetic expansion discussed in our subsequent paper [26].

An interesting observation is that the optimal conditions of different control parameters have roughly the same order of magnitude. For example, at room temperature ($k_B T \approx 200$ cm$^{-1}$) and the experimental relaxation time scale $D^{-1} = 50$ fs ($D = 106 \text{ cm}^{-1}$), the optimal reorganization energy in FMO with zero spatial correlation is $\lambda_{opt} = 92$ cm$^{-1}$ ($\lambda_{opt} = 73$ cm$^{-1}$) for the initial condition at BChl 1 (BChl 6). All these parameters are around in the same energy scale, 100 cm$^{-1}$, which is also close to values of energy difference and electronic coupling between nearest neighbor sites. In our forthcoming paper, based on two asymptotic behaviors of the trapping time in the regimes of weak and strong dissipations, we find that a crude estimation of the optimal dissipation strength is on the same order as site-site coupling [18].

Reliable predictions of the EET process in the light-harvesting systems require an adequate description of dissipative effects. For example, the Forster theory is applicable to incoherent hopping whereas the Redfield equation is applicable to the Markovian limit of exciton dynamics [45, 46]. Due to the nature of classical noise, the Haken-Strobl model is limited to the regime of high temperatures [23]. As derived from the second-order cumulant expansion, the generalized Bloch-Redfield (GBR) equation is shown in this paper to provide a reliable qualitative description of the trapping time in a broad range of dissipation, and successfully predicts the optimization of the ETE with respect to different variables. However, higher order cumulants become relevant for strong/slow dissipation and the corrections to the GBR approach are then necessary. The hierarchical approach with Gaussian noises, applied to FMO [47] and LH2 [51], can be used to improve the quantitative accuracy of our predictions. Alternatively, we can begin with incoherent exicton dynamics as a reference and then include quantum coherence systematically, which can be accomplished by the extension of the noninteracting-blip approximation (NIBA) derived in the two-site systems [52]-[55].

**Acknowledgments**
The work reported here is supported by the National Science Foundation (Grant Number NSF 0806266 and 0556268), the MIT Energy Initiative (MITEI) Seed grant, and the MIT Center for Excitonics funded by DOE (Grant Number DE-SC0001088).